\documentclass{webofc}
\usepackage[varg]{txfonts}   % Web of Conferences font
\begin{document}
\title{Kaonic deuterium and low-energy antikaon-nucleon interaction}

\author{\firstname{Wataru} \lastname{Horiuchi}\inst{1}\fnsep\thanks{\email{whoriuchi@nucl.sci.hokudai.ac.jp}} \and
        \firstname{Tetsuo} \lastname{Hyodo}\inst{2}\fnsep\thanks{\email{hyodo@yukawa.kyoto-u.ac.jp}} \and
        \firstname{Wolfram} \lastname{Weise}\inst{3}\fnsep\thanks{\email{weise@tum.de}}
        % etc.
}

\institute{Department of Physics, Hokkaido University, Sapporo 060-0810, Japan 
\and
          Yukawa Institute for Theoretical Physics, Kyoto University, Kyoto 606-8502, Japan 
\and
           Physics Department, Technical University of Munich, 85748 Garching, Germany
          }

\abstract{%
  A new evaluation of the 1s level shift and width of kaonic deuterium is presented based on an accurate $\bar{K}NN$ three-body calculation, using as input a realistic antikaon-nucleon interaction constrained by the SIDDHARTA kaonic hydrogen data. The three-body Schr\"odinger equation is solved with a superposition of a large number of correlated Gaussian basis functions extending over distance scales up to several hundred fm. The resulting energy shift and width of the kaonic deuterium 1s level are $\Delta E \simeq 0.67$ keV and $\Gamma \simeq 1.02$ keV, with estimated uncertainties at the 10 \% level.
}
\maketitle
\section{Introduction}
\label{intro}
Kaonic hydrogen and kaonic deuterium are the prototype $K^-$-atomic systems to be studied in the quest for constraints on the low-energy antikaon-nucleon interaction near $\bar{K}N$ threshold. Sufficiently accurate data of the strong-interaction energy shift $\Delta E$ and width $\Gamma$ of the 1s level in these kaonic atoms should provide, through their theoretical analysis, key information in order to fix basic $\bar{K}N$ scattering length parameters in both isospin $I = 0,1$ channels. An important step in that direction was taken previously by the SIDDHARTA experiment \cite{siddharta} which determined $\Delta E = 283 \pm 36 \pm 6$ eV and $\Gamma = 541 \pm 89 \pm 22$ eV for kaonic hydrogen. Corresponding data for kaonic deuterium do not yet exist. However, new $K^-d$ measurements are in preparation: SIDDHARTA-2 at LNF \cite{siddharta2} and E57 at J-PARC \cite{e57}. 

\begin{figure}
\centering
\includegraphics[width=7cm]{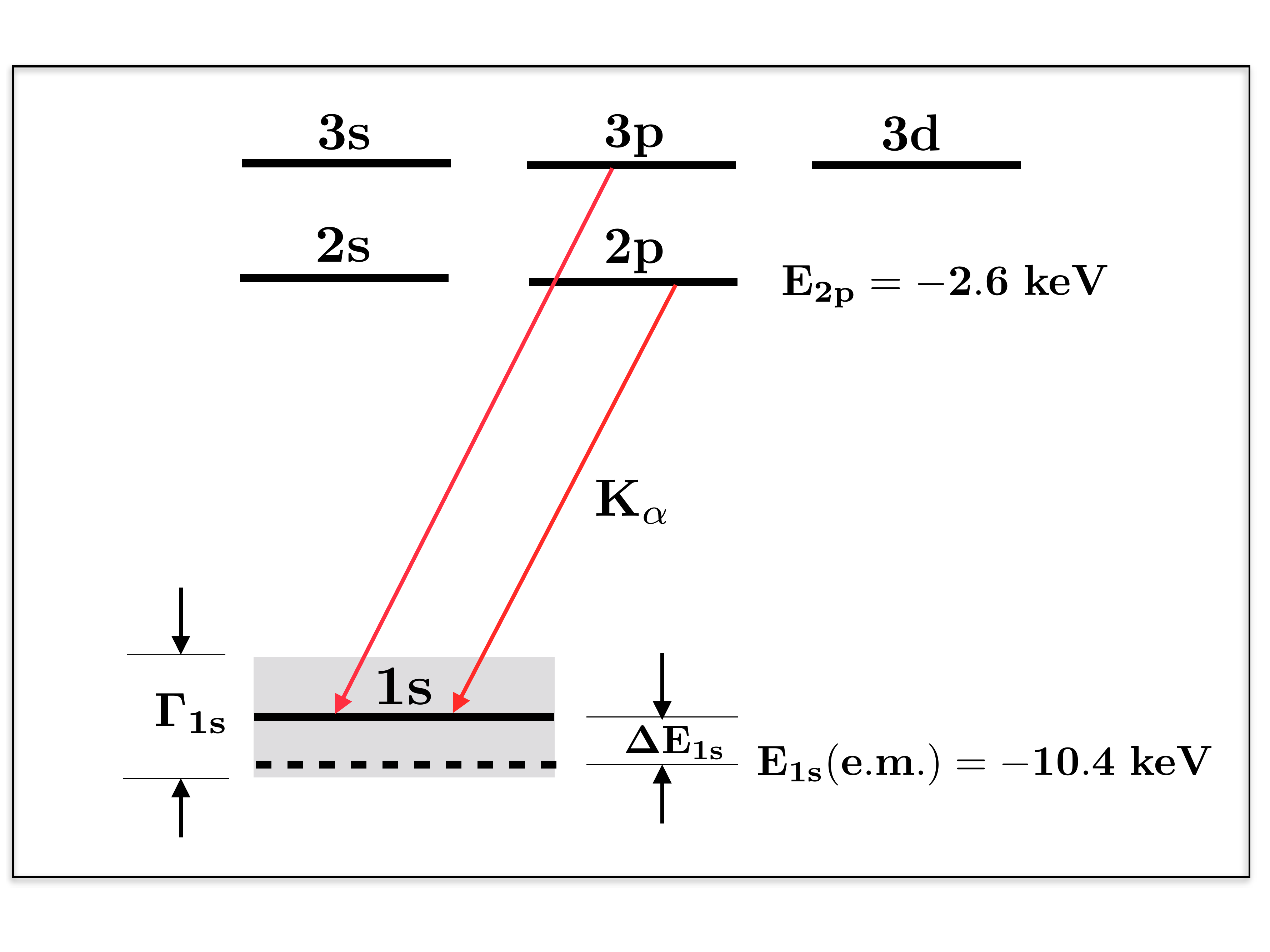}
\caption{Schematic spectrum of kaonic deuterium showing the atomic levels and the $K_\alpha$ transition $(2p\rightarrow 1s)$ together with the strong-interaction shift and width of the 1s level.}
\label{Figure1}       
\end{figure}

These developments call for advanced calculations of the coupled $K^-pn \leftrightarrow \bar{K}^0 nn$ system. The $K^- d$ atomic binding (see the schematic spectrum of Fig.\,\ref{Figure1}) is produced by the attractive $K^- p$ Coulomb interaction. The pertinent atomic length scale is determined by the kaonic deuterium Bohr radius, $a_B(K^-d) \sim 70$ fm. The strong-interaction energy shift and width, on the other hand, result from effects at distance scales of order 1 fm between the nucleons and the antikaon. Clearly, high demands on computational accuracy must be met at {\it all} length scales in order to deal with this challenging three-body problem involving both long-range Coulomb and short-range strong interactions.  

Previous theoretical studies of kaonic deuterium basically fall into two categories. Treatments of the $K^-d$ system as an approximate two-body problem focused on the relationship between kaonic deuterium observables and $\bar{K}N$ two-body scattering lengths \cite{Meissner:2006gx,Gal:2007,Doring:2011xc}. Three-body Faddeev calculations of kaonic deuterium have their own history - see for example Ref.~\cite{Bahaoui:2003}.
Recent advanced Faddeev computations \cite{kdrev,kddol} use separable potentials
constrained by the SIDDHARTA kaonic hydrogen data and have evaluated the $1s$ $K^-d$ atomic state assuming isospin symmetry for the $\bar{K}$ and nucleon doublets. 

The present report gives a brief summary of recent work published in Ref.\,\cite{hohhw:2017}. A novel calculation of kaonic deuterium has been performed solving the full three-body $K^-pn \leftrightarrow \bar{K}^0 nn$ coupled-channels Schr\"odinger equation. A modern $\bar{K}N$ interaction is used as input, namely the Kyoto $\bar{K}N$ potential~\cite{mh} based on chiral SU(3) effective field theory \cite{ihw,Kamiya:2016}. A particular aim of that investigation, apart from achieving highest possible computational precision, has been to set a more stringent constraint for the isospin $I=1$ component of the low-energy $\bar{K}N$ interaction. We recall that kaonic deuterium is expected to be a more sensitive probe than kaonic hydrogen for this $I =1$ channel, given that the ratio of $I=1$ to $I=0$ components is 3:1 in the $K^-d$ system. 

\section{Antikaon-NN three-body problem}
\label{sec-1}

In the present context the theoretical description of kaonic deuterium as a three-body system is defined by the following
coupled-channels Schr\"odinger equation:
 \begin{align}
   \begin{pmatrix}
    \hat{H}_{K^-pn}& \hat{V}_{12}^{\bar{K}N}+\hat{V}_{13}^{\bar{K}N}\\
    \hat{V}_{12}^{\bar{K}N}+\hat{V}_{13}^{\bar{K}N}&\hat{H}_{\bar{K}^0nn}    
   \end{pmatrix}
   \begin{pmatrix}
    \left|K^-pn\right>\\
    \left|\bar{K}^0nn\right>
   \end{pmatrix}=   E
   \begin{pmatrix}
    \left|K^-pn\right>\\
    \left|\bar{K}^0nn\right>
   \end{pmatrix} ~~, 
\label{eq:1}   
 \end{align}
 with the Hamiltonians of the $K^-pn$ and $\bar{K}^0nn$ channels specified as\footnote{Particle indices are introduced  as indicated in Fig.\ref{Figure2}(a): $i=1$ refers to $K^-$ or $\bar{K}^0$ and $i=2,3$ denotes the two nucleons.}
 \begin{align}
   \hat{H}_{K^-pn}&=\sum_{i=1}^3\hat{T}_i-\hat{T}_{\rm cm}+\hat{V}_{23}^{NN}+\sum_{i=2}^3\left(\hat{V}_{1i}^{\bar{K}N}+\hat{V}_{1i}^{\rm EM}\right)~~, \\
   \hat{H}_{\bar{K}^0nn}&=\sum_{i=1}^3\hat{T}_i-\hat{T}_{\rm cm}+\hat{V}_{23}^{NN}+\sum_{i=2}^3\hat{V}_{1i}^{\bar{K}N}+\Delta M~~.
   \end{align}
Here $\hat{T}_i$ are the kinetic energies of the antikaon and the two nucleons. The kinetic energy $\hat{T}_{\rm cm}$ associated with the three-body center-of-mass is properly subtracted as usual. The interaction terms include the nucleon-nucleon potential, $\hat{V}^{NN}$, the electromagnetic interaction $\hat{V}^{EM}$ (i.e. the $K^-p$ Coulomb potential) and the effective antikaon-nucleon potential $\hat{V}^{\bar{K}N}$ that will now be further explained. Also included is the mass difference, $\Delta M$, between the  
$K^-pn$ and $\bar{K}^0nn$ channels. We use physical masses for nucleons and antikaons: $M_p = 938.272$ MeV, $M_n = 939.565$ MeV, $M_{K^-} = 493.677$ MeV and $M_{\bar{K}^0} = 497.648$ MeV, so as to take into account isospin-breaking effects.

\subsection{Antikaon-nucleon effective potential}

\begin{figure*}
\centering
\includegraphics[width=8cm]{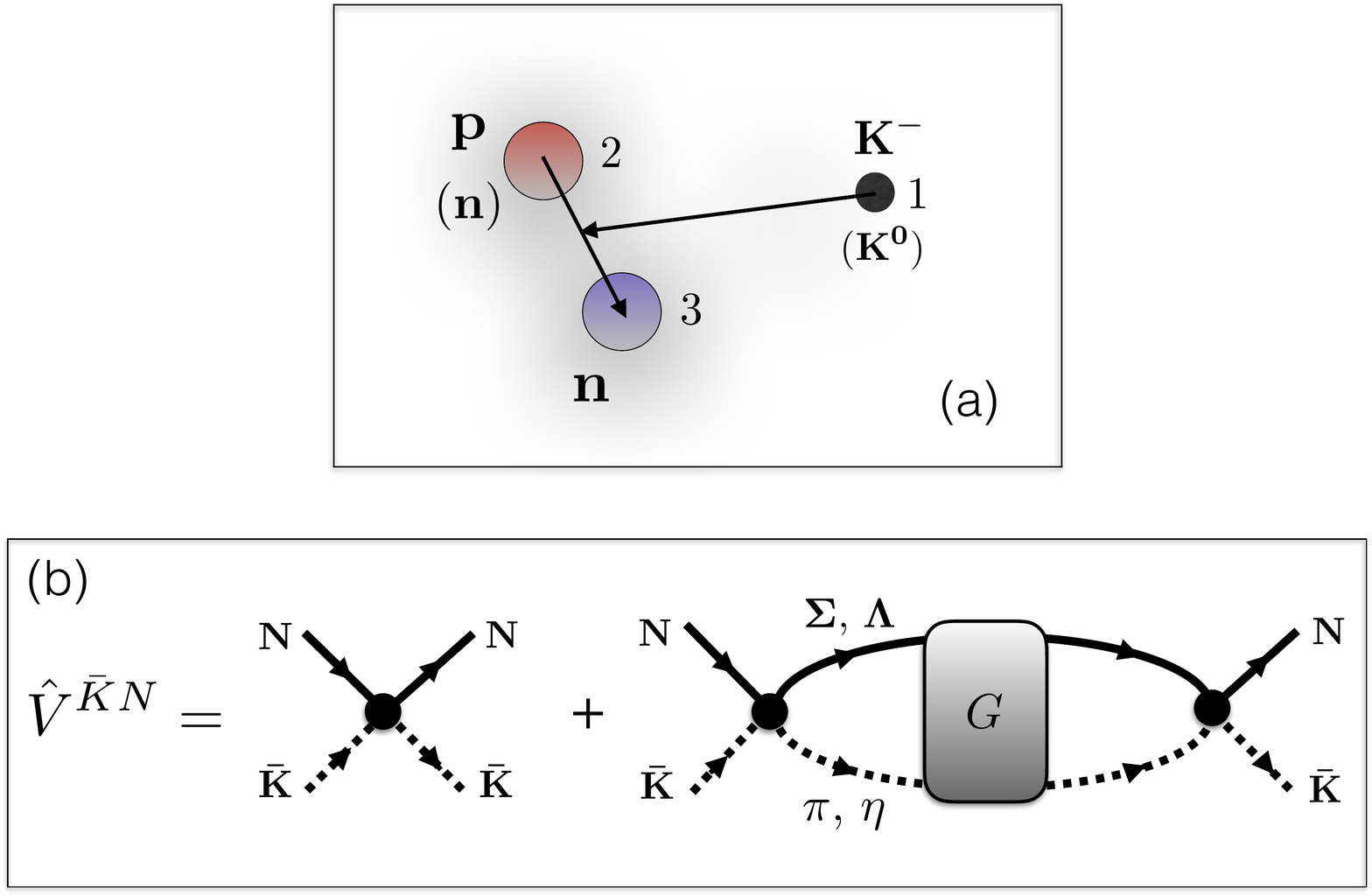}
\caption{(a) Degrees of freedom and notations used in the description of the $K^-pn\leftrightarrow \bar{K}^0 nn$ three-body coupled-channels problem; ~~(b) Construction of the complex effective $\bar{K}N$ potential with inclusion of $\bar{K}N$ couplings to $\pi$- and $\eta$-hyperon $(\Lambda, \Sigma)$ continuum states.}
\label{Figure2}       
\end{figure*}

The present work employs the complex and energy dependent Kyoto $\bar{K}N$ potential \cite{mh}, $\hat{V}^{\bar{K}N}$, as the basic antikaon-nucleon effective interaction. It incorporates important dynamics such as the coupling of the $K^- p$ and $\bar{K}^0 n$ channels to $\pi\Sigma$ and $\pi\Lambda$ continuum states as illustrated in Fig.\,\ref{Figure2}(b). Effects of the decay processes into $\pi\Sigma$ and $\pi\Lambda$ are encoded in the imaginary part of $\hat{V}^{\bar{K}N}$. The $r$-space representation of the potential is parametrised in Gaussian form, $\hat{V}^{\bar{K}N}(r,E) \sim e^{-r^2/b^2}U(E)$, with $b \simeq 0.4$ fm. The strength of the interaction depends on the energy $E \equiv E_{\bar{K}N}$, treated self-consistently in the two-body $\bar{K}N$ Schr\"odinger equation. This potential is based on chiral SU(3) effective field theory \cite{ihw} and constrained by empirical $\bar{K}N$ data. Thanks to systematic improvements with inclusion of higher order terms, $\hat{V}^{\bar{K}N}$ reproduces very well all available low-energy $\bar{K}N$ data ($K^-p$ total and reaction cross sections,
branching ratios at $K^-p$ threshold and the SIDDHARTA data) \cite{ihw}  with an accuracy of $\chi^2/\text{d.o.f.}\simeq
1$. 

The Kyoto $\bar{K}N$ potential in its original form was written in the isospin basis from which the particle basis potential is constructed as
\begin{align}
 \hat{V}^{\bar{K}N}_{ij}=\frac{1}{2}\left[\hat{V}^{\bar{K}N(I=0)}_{ij}+\hat{V}^{\bar{K}N(I=1)}_{ij}\right]
 -\frac{1}{2}\left[\hat{V}^{\bar{K}N(I=0)}_{ij}-\hat{V}^{\bar{K}N(I=1)}_{ij}\right]P_{\tau}^{ij}~~.
\end{align}
The Heisenberg operator $P_{\tau}^{ij}=(1+\tau_i\cdot\tau_j)/2$
exchanges the $i$- and $j$-th particles in the isospin wave function. In the particle basis used throughout this work,  the exchange operator acts as $P_{\tau}^{ij}|K^-n\rangle=|K^-n\rangle$, $P_{\tau}^{ij}|K^-p\rangle=-|\bar{K}^0n\rangle$ and
$P_{\tau}^{ij}|\bar{K}^0n\rangle=-|K^-p\rangle$. The charge exchange channel-coupling between $K^{-}p$ and $\bar{K}^{0}n$ occurs through the isospin dependence of the $\bar{K}N$ interaction.

Before proceeding to the calculation of kaonic deuterium, the input Kyoto effective potential has to be tested in comparison with kaonic hydrogen data and $\bar{K}N$ scattering lengths. We recall that the fitting to the kaonic hydrogen results in Refs.~\cite{ihw} was performed making use of the improved Deser formula~\cite{deser}, whereas in Ref.\,\cite{hohhw:2017} the $\bar{K}N$ coupled-channels Schr\"odinger equation has been solved to evaluate the $K^-$ hydrogen level shift and width. The threshold energy difference between the $K^{-}p$ and $\bar{K^{0}}n$ channels of about 5 MeV must properly be taken into account in the level shift calculation. It is therefore instructive to compare calculations using either physical masses or isospin averaged masses of the $(K^-, \bar{K^0})$ and $(p, n)$ doublets. 
\begin{table}
\caption{Test of the effective $\bar{K}N$ potential $\hat{V}^{\bar{K}N}$ (the Kyoto potential \cite{mh}) for the strong-interaction energy shift and width of kaonic hydrogen. The coupled-channels Schr\"odinger equation (\ref{eq:1}) including the $K^-p$ Coulomb interaction has been solved with a self-consistent determination of the energy-dependence of  $\hat{V}^{\bar{K}N}(E)$.
Values calculated with physical antikaon and nucleon masses (1st row) and isospin-averaged masses (2nd row) are shown in comparison with empirical (SIDDHARTA \cite{siddharta}) results.} 
\label{tab-1}     
\centering
\begin{tabular}{ccc}
\hline
Masses & ~~~$\Delta E$ [eV] & ~~~~$\Gamma$ [eV]  \\\hline
physical & ~~~~~~~283 & ~~~~~~~607 \\
isospin & ~~~~~~~163 & ~~~~~~~574 \\\hline
Exp. \cite{siddharta} & ~~~283 $\pm$ 36 $\pm$ 6 & ~~~541$ \pm$ 89 $\pm$ 22 \\
\hline
\end{tabular}
\end{table}
\begin{table}[tbp]
\caption{Scattering lengths in the $\bar{K}N$ channels from solutions of the coupled-channels Schr\"odinger equation with the Kyoto effective potential $\hat{V}^{\bar{K}N},$ using physical and isospin-averaged masses.}
\centering
\begin{tabular}{cccccc}
\hline
   Masses & ~~~$a_{K^{-}p}$ [fm] & $a_{K^-p - \bar{K}^0n}$ [fm] & $a_{\bar{K}^0 n}$ [fm] & $a_{K^{-}n}$ [fm] \\ \hline
   physical & ~~~$-0.66+i0.89 $ & $-0.85+i0.26 $ & $-0.40+i1.03 $ & $0.58+i0.78 $ \\
   isospin  & ~~~$-0.40+i0.81 $ & $-0.99+i0.04 $ & $-0.40+i0.81 $ & $0.58+i0.77 $  \\
\hline
\end{tabular}
\label{tab-2}
\end{table}

Results are displayed in Tables \ref{tab-1} and \ref{tab-2}. Isospin-breaking mass effects are evidently important in the $\bar{K}N$ threshold physics that generates the 1s energy shift and width of kaonic hydrogen. In particular, the differences between scattering lengths obtained with physical versus isospin-averaged masses translate into a corresponding large difference in $\Delta E_{1s}$. The computation using isospin-averaged masses cannot account for the observed energy shift. 

Concerning the energy dependence of $\hat{V}^{\bar{K}N}(E)$, for kaonic hydrogen the results obtained by setting $E \equiv E_{\bar{K}N} = 0 $ at threshold as input turn out to be equal to those using the self-consistent value of $E$. For kaonic deuterium, determining this energy of the two-body $\bar{K}N$ subsystem within the bound three-body system is a more involved issue that requires further discussion.

\subsection{Solving the three-body Schr\"odinger equation}

The coupled-channels Schr\"odinger equation (\ref{eq:1}) is solved using a variational approach with a large set of basis functions of the generic form
\begin{equation}
\label{eq:wave}
\Phi=\mathcal{A}[\psi^{\rm (space)}\otimes\psi^{\rm (spin)}\otimes\psi^{\rm (isospin)}]~~.
\end{equation}
Correlated Gaussian (CG) basis functions \cite{cgpaper,cgrev} are used to expand the radial part of $\Phi$.
This approach is sufficiently flexible to describe both short- and long-range properties of the
wave function accurately, a necessary condition when dealing with systems such as kaonic deuterium in which very different distance scales must be handled simultaneously. The so-called global vector representation \cite{suzuki08} is applied to describe rotational degrees of freedom, such that the form of the variational wave function does not change under linear transformations of the three-body coordinates. This permits to optimize choices of the Jacobi coordinates for the $K^-pn$ and $\bar{K}^0nn$ channels with their different masses. Details can be found in Ref.\,\cite{hohhw:2017}.

As repeatedly emphasized, short-range strong interactions together with the long-range Coulomb force have to be 
treated with high numerical precision. In order to extract the detailed effects of the $\bar{K}N$ interaction from
the spectrum of kaonic deuterium, its binding energy needs to be calculated with an accuracy of a few eV, a considerable 
computational challenge. The required precision is achieved by expanding the variational wave function (\ref{eq:wave}) in a very large set of basis functions, $\{\Phi_\alpha\}$. Then the generalized eigenvalue problem 
\begin{equation}
\sum_{\beta=1}^N(H_{\alpha\beta}-EB_{\alpha\beta})\,C_\beta=0,
\end{equation}
is solved to determine the coefficients $C_\alpha$ and the energy $E$,
with the Hamiltonian matrix $H_{\alpha\beta}=\langle\Phi_\alpha|H|\Phi_\beta\rangle$ and the overlap matrix
$B_{\alpha\beta}=\langle\Phi_\alpha|\Phi_\beta\rangle$.
Here $N$ is the number of basis functions. To achieve energy convergence for the kaonic atom,
it turns out that basis functions reaching over distance scales from a tenth to several hundreds of fm need to be included. In practice \cite{hohhw:2017}, convergence at the required precision is reached with $N\sim 3000$. The primary basis sizes used extend up to $N\sim 4000$.

In the actual kaonic deuterium calculations, with the energy-dependent Kyoto potential \cite{mh} as input, fixing the two-body energy $E_{\bar{K}N}$ is a non-trivial issue. While the choice $E_{\bar{K}N} = 0$ at the $\bar{K}N$ threshold is justified for kaonic hydrogen, the energy of the $\bar{K}N$ two-body subsystem within the $K^- d$ three-body system is not a well-defined concept. Different prescriptions \cite{dote,barn,Ohnishi17} are available to take into account the motion of the bound nucleons while they interact with the antikaon. In our work \cite{hohhw:2017} we refer to Refs. \cite{dote,Ohnishi17} and start by setting $E = E_{\bar{K}N} = 0$ in the two-body potential $\hat{V}^{\bar{K}N}(E)$ also for kaonic deuterium, the choice we take as default input in the three-body calculations. It is argued in Ref.\,\cite{hohhw:2017} that leading binding corrections to this minimal choice imply a small downward shift in the energy variable, $E \simeq -B/2$, with the deuteron binding energy $B = 2.2$ MeV, in contrast to the much larger shift suggested in Ref.\,\cite{barn}.

\section{Results for kaonic deuterium}

Table \ref{Kdeuterium} lists the energies of the 1s, 2p and 2s states of kaonic deuterium computed by solving the three-body equations (\ref{eq:1}). We recall that the reference energies of the $K^- d$ pure Coulomb two-body system with deuteron point charge are:
\begin{equation}
E_{1s} = -10.406~{\rm keV}~~,~~~~~E_{2p} = E_{2s} = -2.602~{\rm keV}~~.
\end{equation}
In the three-body calculation with Coulomb interaction only, these energies are slightly shifted by the finite extension of the deuteron charge density. With inclusion of the strongly interacting $K^-pn$ and $\bar{K}^0 nn$ coupled channels, i.e. by turning on the effective $\bar{K}N$ potential $\hat{V}^{\bar{K}N}(E = 0)$, the $1s$ level experiences a complex energy shift while the $2p$ level remains unchanged. As expected the strong-interaction shift of the $1s$ energy can be directly determined by measuring the $2p\rightarrow 1s$ transition (the $K_\alpha$ line). In summary, our predicted kaonic deuterium $1s$ level shift is (up to corrections to be discussed below):
\begin{equation}
\Delta E-i\frac{\Gamma}{2}= (670-i\,508)~\text{eV}~~,
\end{equation}
using the Kyoto $\bar{K}N$ potential.
These values are roughly consistent with those found in a recent Faddeev calculation~\cite{kdrev},
although the basic interactions used there are different from ours.

\begin{table*}[tbp]
\caption{Calculated energy spectrum of kaonic deuterium. Results of the three-body calculation are shown 
with Coulomb interaction only (upper row) and with inclusion of the strong $\bar{K}N$ interaction (lower row). Energies of the $1s, 2p$ and $2s$ states are given relative to the $K^-d$ threshold, $m_{K^-} + M_d$, with the deuteron mass $M_d = 1875.613$ MeV.}
\centering
\begin{tabular}{cccccc} 
\hline
                &$E_{1s}~$[keV] & $E_{2p}~$[keV] & $E_{2s}~$[keV] \\ 
\hline
Coulomb       & $-$10.398 & $-$2.602& $-$2.600\\
Coulomb$+\bar{K}N$&$-$9.736$-i\,$0.508& $-$2.602$-i\,$0.000&$-$2.517$-i\,$0.067\\ 
\hline
\end{tabular}
\label{Kdeuterium}
\end{table*}

The results just outlined have been obtained using physical masses for the antikaons and for proton and neutron. Notably, a calculation with isospin-averaged masses yields a $1s$ energy shift and width that differ by only a few eV from the previously mentioned values. Unlike the situation in kaonic hydrogen, effects of isospin breaking turn out to be small in kaonic deuterium. This feature is further discussed and explained in Ref.\,\cite{hohhw:2017}.

Up to this point the determination of the width $\Gamma$ incorporates the decay channels $\bar{K}N\rightarrow\pi Y$, where $Y$ stands for $\Lambda$ and $\Sigma$ hyperons. The question arises about possible additional contributions to the width from antikaon absorption on two nucleons, with the coupled $K^-pn$ and $\bar{K}^0 nn$ channels decaying into $\Lambda n + \Sigma^0 n + \Sigma^- p$. Early measurements at Brookhaven with $K^-$ stopped on liquid deuterium in the BNL bubble chamber~\cite{Veirs:1970fs} demonstrated that these processes are strongly suppressed as compared to the leading single-nucleon channels, $\bar{K}N\rightarrow \pi Y$. The ratio of two-nucleon absorption reactions to the single-nucleon processes was found to be as small as $(1.2\pm 0.1)\%$~\cite{Veirs:1970fs}. Taking this value for orientation, the kaonic deuterium $1s$ width would increase through two-nucleon absorption by only about $10$ eV, a correction that can be safely neglected within an estimated uncertainty range of approximately $10\,\%$ assigned to the calculated width of about a keV. The smallness of the two-body absorptive width can be understood as follows. Kinematical conditions for the $\bar{K}NN\rightarrow YN$ process require a large momentum transfer of order 1 GeV/c to be provided by the initial deuteron wave function at short distances. The probability is low for this to take place in such a weakly bound, dilute system. 

A further source of possible uncertainties is related to the energy dependence of the $\bar{K}N$ potential, $\hat{V}^{\bar{K}N}(E_{\bar{K}N})$. As mentioned, $E_{\bar{K}N} = 0$ at threshold was set in the present calculations. The binding of the nucleons in the deuteron may cause a shift of $E_{\bar{K}N}$ towards the subthreshold region. Our estimate, derived and discussed in the Appendix of Ref.\,\cite{hohhw:2017}, suggests a small average shift, $E_{\bar{K}N} =-B_{d}/2\sim -1.1$ MeV, involving the deuteron binding energy $B_{d}$. 
The changes thus induced by correcting the energy dependence in $\hat{V}^{\bar{K}N}(E_{\bar{K}N})$ for deuteron binding tend to increase the width of kaonic deuterium by about 10\%, while the corresponding energy shift changes by only about 5\%. Nonetheless, a more detailed treatment of such kinematical corrections within the $\bar{K}NN$ three-body problem is of some importance, given the strong subthreshold energy dependence of the $K^-p$ scattering amplitude involving the $\Lambda(1405)$.

\section{Test of improved Deser formulae for kaonic deuterium}

The improved Deser formula~\cite{deser,Meissner:2006gx},
derived from non-relativistic effective field theory (EFT),
is frequently used in the investigation of strong-interaction effects in hadronic atoms.
This relation connects the $1s$ level shift $\Delta E$ and width $\Gamma$ of a kaonic atom with the $K^-$-nucleus scattering length, $a$, as follows:
\begin{equation}
 \Delta E - \frac{i\Gamma}{2}=-2\mu^2\alpha^3 a\,[1-2\mu\alpha(\ln
 \alpha-\,a],\label{imp_deser}
\end{equation}
where $\mu$ is the kaon-nucleus reduced mass and $\alpha$ is the fine
structure constant. 
The logarithmically enhanced correction term can be resummed to all orders~\cite{Baru:2009tx}, providing a ``double-improved" Deser formula: 
\begin{equation}
 \Delta E - \frac{i\Gamma}{2}=-\frac{2\mu^2\alpha^3a}{1+2\mu\alpha(\ln
 \alpha-1)\,a}.\label{imp_deser2}
\end{equation}

A first check for kaonic hydrogen (see Table\,\ref{tbl:Kpresult}) using the calculated $K^-p$ scattering length of Table\,\ref{tab-2} points out that the modified Deser formula (\ref{imp_deser}) works well in this case, with a deviation of just about 10 eV from the full solution of the $\bar{K}N$ Schr\"odinger equation. The resummed version (\ref{imp_deser2}) improves the comparison further, down to a difference of only 2 eV.  
\begin{table}[tbp]
\caption{Strong-interaction energy shift and width of the kaonic hydrogen $1s$ state, obtained by solving the two-body Schr\"odinger equation with the Kyoto $\bar{K}N$ potential, and by using the improved Deser formula and its resummed version.} 
\begin{center}
\begin{tabular}{lcc}
\hline
   Kaonic hydrogen& $\Delta E$ ~[eV] & $\Gamma$~ [eV] \\
   \hline
   Full Schr\"odinger equation & 283 & 607  \\
   Improved Deser formula~\eqref{imp_deser} & 293 & 596  \\
   Resummed formula~\eqref{imp_deser2} & 284 & 605 \\
\hline
\end{tabular}
\end{center}
\label{tbl:Kpresult}
\end{table}%

Next we compare the results of the full three-body calculation for kaonic deuterium with the estimates derived from Eqs.~(\ref{imp_deser}) and (\ref{imp_deser2}). This requires as input the $K^{-}d$ scattering length, $a_{K^{-}d}$. In the fixed center approximation (FCA) for the nucleons,
$a_{K^-d}$ derived from a multiple scattering series is given as \cite{Kamalov:2000iy,Meissner:2006gx}
\begin{align}
a_{K^-d}&=\frac{\mu_{K^- d}}{m_{K^-}}\int d^3r\, \rho_d(r)\,\tilde a_{K^-d}(r)~~,\label{rusetsky}\\
\tilde a_{K^-d}(r)&=\frac{\tilde a_{p}+\tilde a_{n}
+(2\tilde a_{p}\tilde a_{n}-\tilde a_{\rm ex}^2)/r
-2\tilde a_{\rm ex}^2\tilde
a_{n}/r^2}
{1-\tilde a_{p}\tilde a_{n}/r^2+\tilde
a_{\rm ex}^2\tilde a_{n}/r^3}~~,
\label{eq:akd}
\end{align}
with the $K^-$-deuteron reduced mass $\mu_{K^{-}d}$, and $\rho_d(r)$ is the nucleon density distribution in the deuteron. The lab-frame scattering lengths are denoted
$\tilde{a}_{p}\equiv\tilde{a}_{K^{-}p}$, $\tilde{a}_{n}\equiv\tilde{a}_{K^{-}n}$, and $\tilde a_{\rm ex}^2\equiv
\tilde a_{K^-p\text{-}\bar{K}^0n}^2/(1+\tilde a_{\bar{K}^0n}/r)$ refers to $K^-p\leftrightarrow \bar{K}^0 n$ charge exchange within the multiple scattering chain. The lab frame quantities are related to the standard scattering lengths in the c.m. frame by 
$\tilde a_{\bar{K}N}\equiv \frac{m_K}{\mu_{\bar{K}N}}a_{\bar{K}N}$
with the $\bar{K}N$ reduced mass $\mu_{\bar{K}N}$. With the set of scattering lengths calculated using the Kyoto $\bar{K}N$ potential and listed in Table\,\ref{tab-2}, the $K^{-}d$ scattering length derived from Eqs.\,(\ref{rusetsky}) and (\ref{eq:akd}) is then:
\begin{equation}
 a_{K^-d}= (-1.42 + i \,1.60)\,\text{fm}~~.
 \label{eq:aKd}
\end{equation}

\begin{table}[tbp]
\caption{Strong-interaction energy shift and width of the kaonic deuterium $1s$ state, obtained by solving the three-body Schr\"odinger equation with the Kyoto $\bar{K}N$ potential, and by using the improved Deser formula and its resummed version.}
\begin{center}
\begin{tabular}{lcc}
\hline
   Kaonic deuterium& $\Delta E$ ~[eV] & $\Gamma$~ [eV]  \\
   \hline
   Full Schr\"odinger equation & 670 & 1016  \\
   Improved Deser formula~\eqref{imp_deser} & 910 & 989  \\
   Resummed formula~\eqref{imp_deser2} & 818 & 1188 \\
\hline
\end{tabular}
\end{center}
\label{tbl:Kdresult}
\end{table}%
Using this value of $a_{K^- d}$ we apply the improved Deser formulae to kaonic deuterium. The results are summarized in Table~\ref{tbl:Kdresult} together with those from the full three-body calculation. Unlike the kaonic hydrogen case, it is evident that estimates based on both versions, Eqs.~\eqref{imp_deser} and \eqref{imp_deser2}, of the Deser formulae are not reliable for kaonic deuterium. In particular, the deviations in the energy shift $\Delta E$ from the full three-body calculation are significantly larger than 100 eV. Of course the $K^{-}d$ scattering length \eqref{eq:aKd} is estimated in the FCA limit, hence it is expected to differ from the "exact" $a_{K^-d}$. The importance of nucleon recoil corrections, naturally included in the full three-body calculation but neglected in FCA, is discussed in Refs.~\cite{Baru:2009tx,Mai:2014uma}. 

\section{Summary and conclusions}

\begin{itemize}
\item{Accurate three-body variational calculations have been performed for the spectrum of
kaonic deuterium and the evaluation of the $1s$ level shift and width.
The $\bar{K}NN$ three-body wave function is expressed as a superposition of a large set of
correlated Gaussian basis functions. The computational challenge, namely a high-precision treatment of
both short-range strong interactions and long-range Coulomb force, requires a very large model space 
covering all distance scales from 0.1 fm to several hundreds of fm.}

\item{The $\bar{K}N$ strong interaction is treated in terms of a complex effective potential that accurately reproduces previous results of coupled-channels calculations based on chiral $SU(3)$ dynamics.  Among the computed energies of $1s$, $2s$ and $2p$ states in kaonic deuterium the $\bar{K}N$ strong interaction affects only the $s$ states,
inducing energy shifts from the levels characteristic of the pure Coulomb and point charge limit
of the $K^-d$ atomic system. No energy shift is found for the $2p$ state, so that
the $1s$ level shift can be directly associated with the transition energy from the $2p$ to the $1s$ state. The calculated $1s$ level shift and width of kaonic deuterium is
$\Delta E-i\Gamma/2=(670-i\,508)$ eV, corresponding to a $2p \rightarrow 1s$ transition energy of $7.134$ keV.
Following discussions in Ref.\,\cite{hohhw:2017} we assign uncertainties of about $10\%$  to $\Gamma$ and less than $10\%$ to $\Delta E$ (not counting the approximately $20\%$ uncertainties in the empirical SIDDHARTA kaonic hydrogen constraints).} 

\item{In view of upcoming experimental investigations we have also performed a detailed test of the sensitivity of kaonic deuterium observables with respect to the $I=1$ component in the $\bar{K}N$ interaction, by varying selectively
the real part of the $I=1$ $\bar{K}N$ potential strength within the uncertainty limits deduced from the kaonic hydrogen data. One can conclude from this test that the $1s$ level shift of kaonic deuterium is indeed expected to provide a significantly improved constraint on the $I=1$ component, as compared to the SIDDHARTA kaonic hydrogen measurement~\cite{siddharta}, if the kaonic deuterium level shift can be determined within $\sim 25\%$ accuracy (corresponding to $\sim 2\%$ in the $2p\rightarrow1s$ transition energy). This sets the physics focus on the yet basically unknown $K^-$neutron sector of the low-energy $\bar{K}N$ interaction.}
\end{itemize}

\subsection{Acknowledgments}

We thank Tsubasa Hoshino and Shota Ohnishi for their crucial contributions to the computational tasks of this work. The numerical calculations were in part performed on the supercomputer (CRAY XC40) at the Yukawa Institute for Theoretical Physics (YITP), Kyoto University.
This work is in part supported by the Grants-in-Aid for Scientific Research on Innovative Areas from MEXT (Grant No. 2404:24105008), by JSPS KAKENHI Grant No. JP16K17694, by the Yukawa International Program for Quark-Hadron Sciences (YIPQS), and by the DFG Cluster of Excellence {\it Origin and Structure of the Universe}. One of the authors (W. W.) gratefully acknowledges the hospitality of YITP and Kyoto University during his three-months visit in 2017.

%\newpage
%

\end{document}